\documentclass[iop]{emulateapj}
\usepackage{amsmath}
\usepackage{longtable}
\usepackage{threeparttablex} 

\shorttitle{USP Mutual Inclinations}
\shortauthors{Li et al.}

\begin{document}

\title{Mutual Inclination Excitation by Stellar Oblateness}
\author{Gongjie Li \altaffilmark{1}, Fei Dai \altaffilmark{2}, Juliette Becker \altaffilmark{2}}
\affil{$^1$ Center for Relativistic Astrophysics, School of Physics, Georgia Institute of Technology, Atlanta, GA 30332, USA}
\email{gongjie.li@physics.gatech.edu}
\affil{$^2$ Division of Geological and Planetary Sciences, California Institute of Technology, Pasadena, CA 91125, USA}

\begin{abstract}
Ultra-short-period planets (USPs) provide important clues to planetary formation and migration. Recently, it is found that the mutual inclinations of the planetary systems are larger if the inner orbits are closer ($\lesssim 5R_*$) and if the planetary period ratios are larger ($P_2/P_1 \gtrsim 5$) \citep{Dai18}. This suggests that the USPs experienced both inclination excitation and orbital shrinkage. Here we investigate the increase in the mutual inclination due to stellar oblateness. We find that the stellar oblateness (within $\sim 1$Gyr) is sufficient to enhance the mutual inclination to explain the observed signatures. This suggests that the USPs can migrate closer to the host star in a near coplanar configuration with their planetary companions (e.g., disk migration+tides or in-situ+tides), before mutual inclination gets excited due to stellar oblateness. 
\bigskip
\end{abstract}


\section{Introduction}
The ultra-short-period planets (USPs) can be loosely defined as terrestrial or super-Earth planets ($<2R_\oplus$) with orbital periods $<1$ day and an occurrence of 0.5\% around sun-like stars \citep{SanchisOjeda14}. The mere presence of these extremely irradiated worlds are puzzling to traditional planet formation theory: their current-day orbits are well within the dust sublimation radius. It is hard to imagine how {\it in-situ} core accretion could have worked without solid materials. Two classes of formation scenarios have been proposed to explain the USP formation. Both require USPs to initially form on wider orbit. One model posits that after accreting on a more distant orbit, these planets experience slow tidal decay and largely remain on circular orbits \citep[e.g., in-situ+tides by][]{Lee17}. On the other hand, these planets can also migrate by a more violent channel through secular interactions with their planetary companions, which launch USPs into eccentric and inclined orbits before tide circularizes and shrinks the orbits \citep{Petrovich19,Pu19}. 

\citet{Dai18} showed empirically that the observed USPs do tend to have a larger mutual inclination (6.7$^\circ$) than longer-period {\it Kepler} multi-planet systems (2.0$^\circ$). Moreover, they showed that a large mutual inclination is also associated with a larger orbital period ratio between USP and its neighboring planets. It has been shown that dynamical hotter formation scenarios can generate orbital migration and mutual inclination of USPs simultaneously, and explain the observed signatures \citep{Petrovich19, Pu19}.

In this work, we will revisit the formation scenario through disk migration, which initially leads to near coplanar planetary configurations. We examine the plausibility that the stellar quadrupole potential leads to higher mutual inclinations after planets migrate to their current locations \citep{Spalding16}. In this case, the quadrupole moment of a host star may induce the nodal precession of a short-period planet. If the whole system starts with a small but non-zero stellar obliquity $\beta$, the differential precession frequencies of the various planets enhance the orbital inclinations. An observer will see a mutual inclination up to $2\beta$. 

Our solar system has a small stellar obliquity of $\sim 7^\circ$, but the obliquities of the exoplanetary systems have a wide range \citep[e.g.,][]{Hebrard08, Winn09, Munoz18}. Many theories have been proposed to explain these spin-orbit misalignments. For instance, the planetary orbit can be tilted due to planetary interactions \citep[e.g.,][]{Fabrycky07, Naoz11, Wu11, Li14}. In addition, the primordial stellar obliquity can be produced by magnetic interaction between the star and disk \citep[e.g.,][]{Lai11, Spalding15}, tilt of the disk due to stellar companions \citep{Batygin12, Zanazzi18} and fluid dynamical effects inside the stars \citep{Rogers12}. This paper operates within the early excitation framework. We show that the primordial stellar obliquity can also explain the larger mutual inclinations of the USPs reported by \citet{Dai18} without the need to resort to secular interactions between planets. 

The article is organized as the following: in section \textsection \ref{s:examples}, we illustrate the evolution of planetary mutual inclination due to stellar oblateness, and in section \textsection \ref{s:emp}, we estimate the stellar oblateness using the observed stellar rotation rate. Then, we simulate the planetary mutual inclination and compare that with the observation in section \textsection \ref{s:obs}, and finally in section \textsection \ref{s:dis}, we discuss possible formation scenarios of the USPs based on our results.

\section{Mutual Inclination Variations Due to J-2 Precession}
\label{s:examples}
Planetary orbital plane can precess due to the $J_2$ potential of a oblate star. When the planet-planet coupling is weaker than the planet-star coupling, the planetary orbit precess at different rates, and this could lead to increases in the planetary mutual inclinations. Without spin-orbit resonances, the maximum mutual inclination could reach $\sim$ twice that of the planet's spin-orbit misalignment \citep{Spalding16}. To illustrate how the mutual inclinations vary due to stellar oblateness, we use Kepler-342 as an example. Kepler-342 contains a star of radius $1.5$R$_{\odot}$ with four planets of radii around $0.89$R$_{\oplus}$, $2.3$R$_{\oplus}$, $2.0$R$_{\oplus}$, $2.5$R$_{\oplus}$, and periods of $1.64$ days, $15.17$ days, $26.23$ days and $39.46$ days \citep{Rowe14, Morton16}. We choose Kepler-342 arbitrarily.

\begin{figure}[h]
\center
    \includegraphics[width=0.4\textwidth]{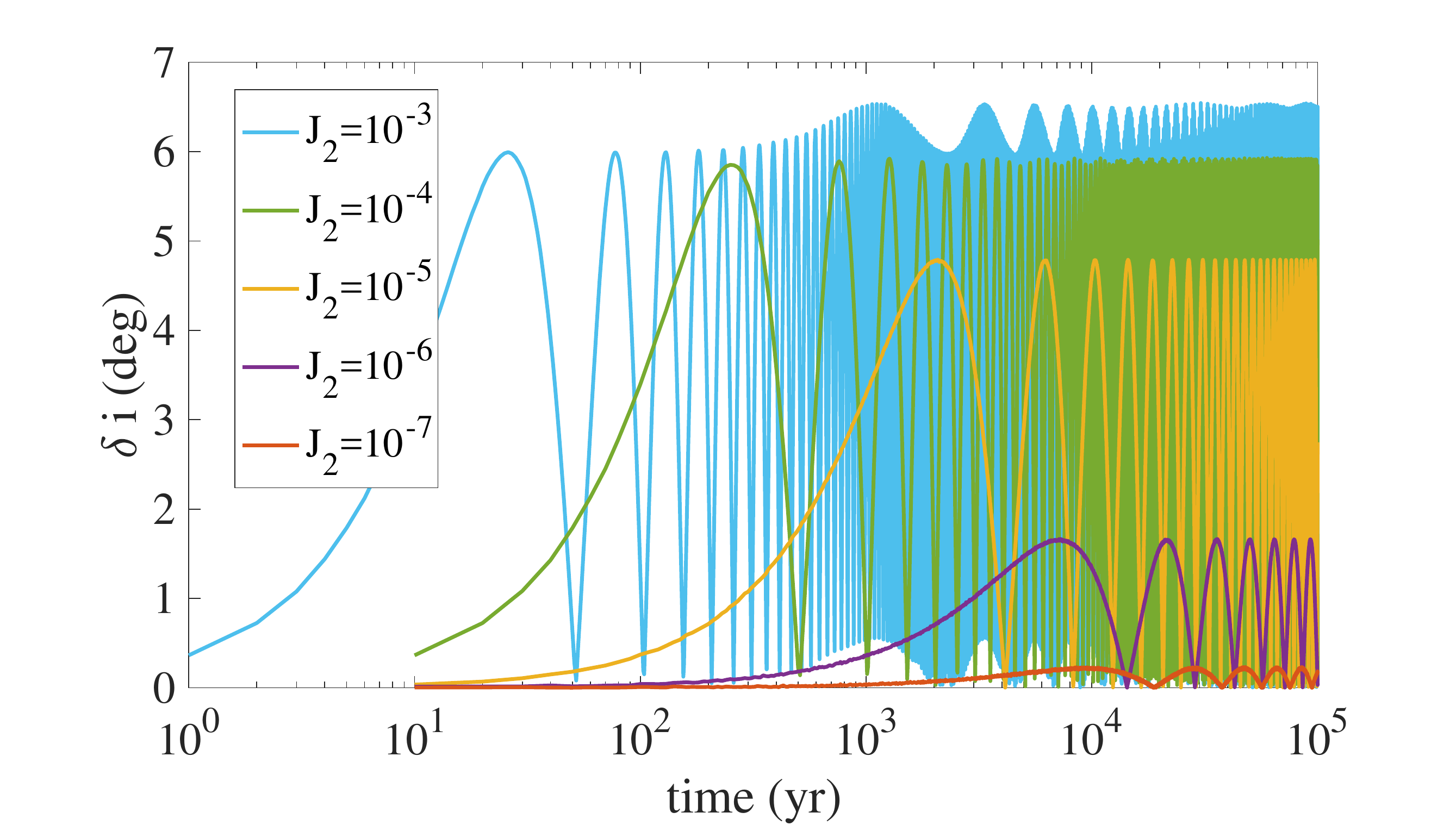}
    \caption{The mutual inclination evolution for Kepler-342 with different stellar $J_2$. Mutual inclination increases as $J_2$ increases, and the maximum mutual inclination reaches around twice stellar obliquity when $J_2$ reaches above $\sim10^{-4}$. Larger $J_2$ potential also leads to faster mutual inclination oscillations\label{f:K342}}
\end{figure}

We use {\it Mercury} N-body simulation package to evolve the system of Kepler-342, and we include $J_2$ precession and $1PN$ correction for general relativistic effects in the user-defined force subroutine \citep{Blanchet06}. The effects of the $1PN$ GR precession is negligible in the mutual inclination, since it only leads to precession in the argument of pericenter. The higher $PN$ order precessions are much weaker comparing with that of the $J_2$ potential here, so we neglect them in the simulations. We obtain the masses of the planets using the empirical fit of the mass-radius relation by \citet{Chen17}. For illustration, we set the planets to be in the same plane, inclined relative to the spin-axis of the star by $3^\circ$. We calculated the angular momentum vector for each of the planetary orbit and then estimated the maximum mutual inclination ($\delta i$) between the planets as shown in Figure \ref{f:K342}. 

To estimate the dominance of the planetary interactions, we calculated the secular nodal oscillation frequency and compare that with the $J_2$ precession. The $J_2$ precession frequency can be estimated as the following
\begin{equation}
    f_{J2}=\frac{3}{2}n\times J_2\Big(\frac{R_*}{a}\Big)^2 ,
\end{equation}
where $n$ is the planetary orbital frequency, $R_*$ is the stellar radius, and $a$ is the semi-major axis. $J_2$ precession frequency is sensitive to the planetary semi-major axis ($f_{J2} \propto a^{-3.5}$), so the differential precession frequency due to $J_2$ is dominated by that of the innermost planet. Thus, we compute the $J_2$ precession frequency of the innermost planet, and compare to that due to planetary interactions.

To estimate the precession frequencies caused by companion planets, we obtain the secular nodal oscillation frequencies using the Lagrange-Laplace method. We do not take the approximation of small semi-major axis ratios, since our sample consists of systems with large period ratios as well as those that are more compact. Generally, the planets couple with each other and are affected by a number of eigenmodes. Then, mutual inclinations can be excited when the $J_2$ precession frequency becomes larger than the slowest modal frequency that affects the innermost planets non-negligibly. Note that the $J_2$ precession does not need to be faster than all the modal frequencies. The exact amplitudes of the modes depend on the boundary condition (inclinations of the planets at an epoch), for simplicity we select the slowest modal frequency requiring the eigenvector component of the innermost planet is larger than $0.1$.

For Kepler-342, the slowest inclination oscillation modal frequency is  $f_{LL}=66''/yr$, corresponding to the same $J_2$ precession frequency ($f_{J2}$) when $J_2=5\times10^{-5}$. Thus, we could see the maximum mutual inclination reaches $\sim 6^\circ$ (twice of the spin-orbit misalignment) for $J_2 \gtrsim 10^{-4}$, and the mutual inclination decreases as $J_2$ decreases below $\sim 10^{-5}$ in Figure \ref{f:K342}.

\begin{figure}[h]
\center
    \includegraphics[width=0.4\textwidth]{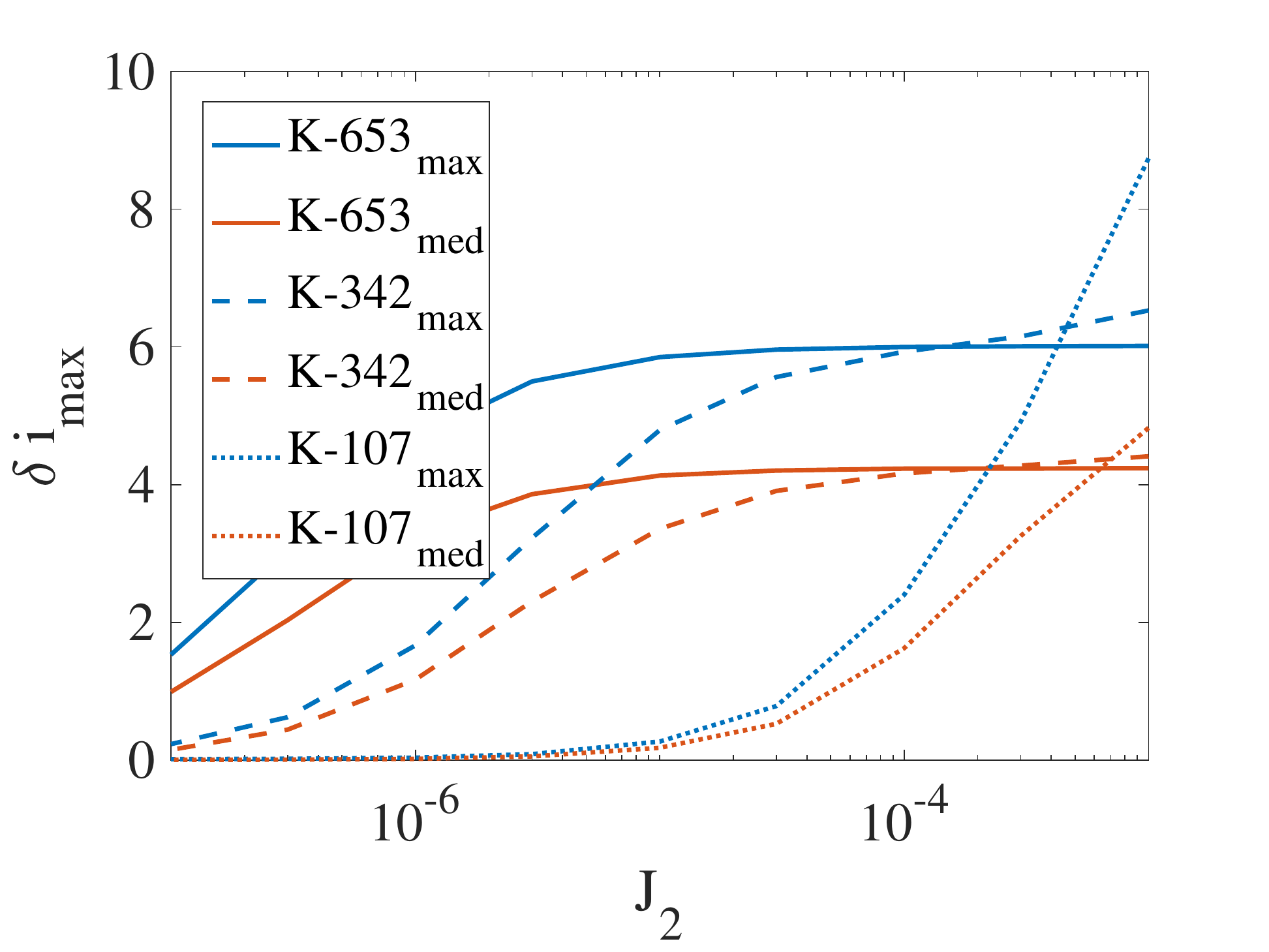}
    \includegraphics[width=0.4\textwidth]{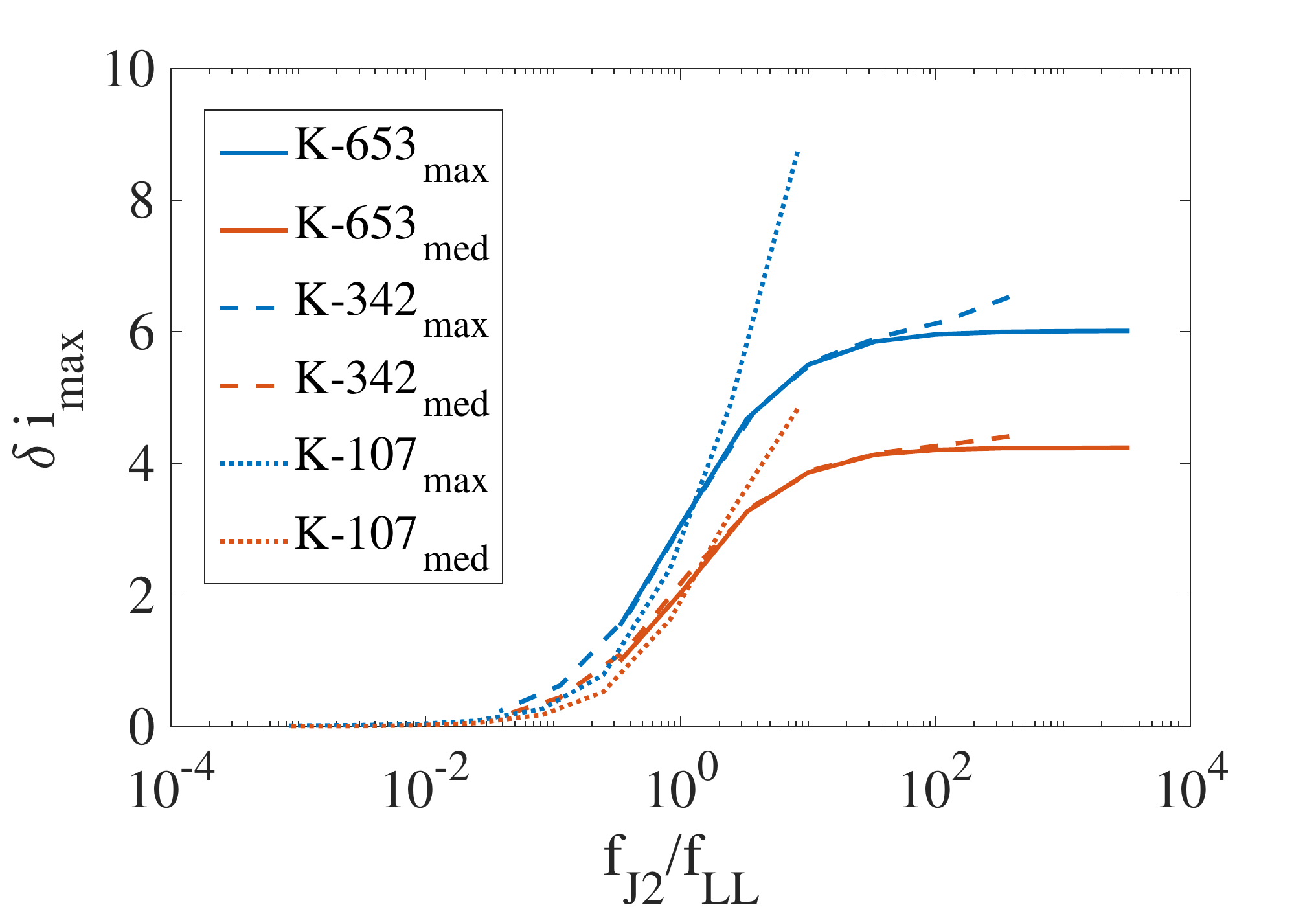}
    \caption{The maximum mutual inclination between the planets as a function of stellar $J_2$ (top) and as a function of $f_{J2}/f_{LL}$ (bottom), where $f_{J2}$ is the nodal precession frequency due to the stellar oblateness, and $f_{LL}$ is the nodal oscillation frequency due to the companion planets (calculated using the Lagrange-Laplace method). When the two frequencies match ($f_{J2} \sim f_{LL}$), the mutual inclination between the planets becomes close to the stellar obliquity. The mutual inclination as a function of the frequency ratio shows some generic features for less compact planetary systems.}
    \label{f:frat}
\end{figure}

The mutual inclination depends on the planetary configuration and the stellar oblateness (\citealt{Spalding16}, Becker et al. in prep). When the innermost planet is closer to the host star, the effects of $J_2$ precession is stronger, and when the planets are closer to each other, planetary perturbations are stronger. However, the dependence of the mutual inclination on the ratio between $f_{LL}$ and $f_{J2}$ is generic for different planetary systems. For illustration, we include three systems with different inner planetary period ratios: Kepler-107 ($P_2/P_1 = 1.54$), Kepler-342 ($P_2/P_1 = 9.23$), and Kepler-653 ($P_2/P_1 = 16.34$) in Figure \ref{f:frat}. 

The top panel of Figure \ref{f:frat} shows the maximum (blue) and medium (red) mutual inclination versus stellar $J_2$ moment. The maximum and medium inclinations are obtained by integrating the systems for $10^5$ yrs (longer than the maximum mutual inclination oscillation period). It shows that the mutual inclination remains low when the stellar oblateness is low, and more compact planetary systems require a larger stellar $J_2$ moment to excite the mutual inclination. 

The bottom panel of Figure \ref{f:frat} shows the maximum mutual inclination vs. the frequency ratio. Although the mutual inclination excitation depends on the architectural properties of the systems, the general behavior versus the frequency ratio can be quite similar for the less compact systems (e.g., Kepler-342 and Kepler-653), where the planetary couplings are weak. For compact systems (e.g., Kepler-107 or Kepler-11 as discussed in \citealt{Spalding16}), the mutual inclination can be further enhanced due to planet-planet interactions, allowing it to reach above twice of the spin-orbit misalignment. More extremely, the mutual inclination could lead to instability of the compact system as discussed in \citet{Spalding16}. For less compact configurations, the systems are quite stable including the $J_2$ precession. In particular, Kepler-342 and Kepler-653 are both stable for at least $10$Myr based on our simulations even with a large $J_2$ of $10^{-3}$. Note that we fix $J_2$ to be a constant in our simulations for simplicity, since involving a time dependent $J_2$ would not change our results qualitatively (see detailed discussions with an evolving $J_2$ in Becker et al. in prep).

\section{Empirical $J_2$ of the sample}
\label{s:emp}
The oblateness of the stars depends on the stellar rotation rate ($J_2\propto\Omega^2$), which decays over time roughly following the Skumanich relation ($\Omega\propto1/\sqrt{t}$). Thus, $J_2$ decays linearly with time. To estimate the effects of the $J_2$ precession, we first calculate the current $J_2$ of the stars in our sample as a reference on the order of $J_2$ in the past when USPs migrated to their current locations.

To estimate the $J_2$ of the stars in our sample we use the Equations below \citep[e.g.,][]{Sterne39, Spalding16}: 
\begin{equation}
J_2 = 1/3(\Omega/\Omega_b)^2 k_2 ,
\end{equation}
where $\Omega$ is the stellar rotational frequency;  $\Omega_b$ is the rotational frequency at break-up; $k_2$ is the love number. $\Omega_b$ is further related to the host star mass and radius:
\begin{equation}
T_b = 2\pi/\Omega_b \approx 1/3 (M_\star/M_\odot)^{-1/2} (R_\star/2R_\odot)^{3/2}~{\rm days} ,
\end{equation}
and we set $k_2 = 0.014$ for all the stars, assuming a $n=3$ polytrope \citep{Sterne39, Yip}. Note that this is only a rough estimate. For a comparison, $k_{2,\odot} \sim 0.03$ and $J_{2,\odot}$ is estimated as $2\times 10^{-7}$ based on helioseismology and solar interior models \citep[e.g.,][]{Mecheri04}, and substituting the solar values into these equations gives $10^{-7}$.

We measured the stellar rotation periods of the stars in our sample using the quasi-sinusoidal flux variations in the {\it Kepler} light curves. We used both auto-correlation function \citep{McQuillan} and Lomb-Scargle periodogram \citep{Lomb,Scargle}. We considered a rotation period a solid detection when the auto-correlation function and the periodogram gave consistent results. We also compared our results with \citet{McQuillan} and \citet{Angus}. The results are presented in Table \ref{tab:Prot}.

To estimate the $J_2$ of the star, we need the stellar masses and radii. We cross matched our star list with the California-Kepler-Survey. If no match was found, we adopted the stellar parameters reported by \citet{Mathur}.

Many of the stars in our sample do not have precise age estimates. Thus, for comparison, we estimated the $J_2$ of stars in three young clusters with good age estimates: Pleiades \citep{Rebull}, Praesepe \citep{Douglas} and NGC 6811 \citep{Curtis}. We used the stellar rotation periods reported by these works. We estimate the stellar masses and radii using the reported effective temperature and the empirical relationship in \citet{Boyajian} and \citet{Gonzales}. We did not include any younger clusters because younger stars whose more inflated radii likely affected the estimated $J_2$. 

\begin{figure*}[h]
 \includegraphics[width=1.\textwidth]{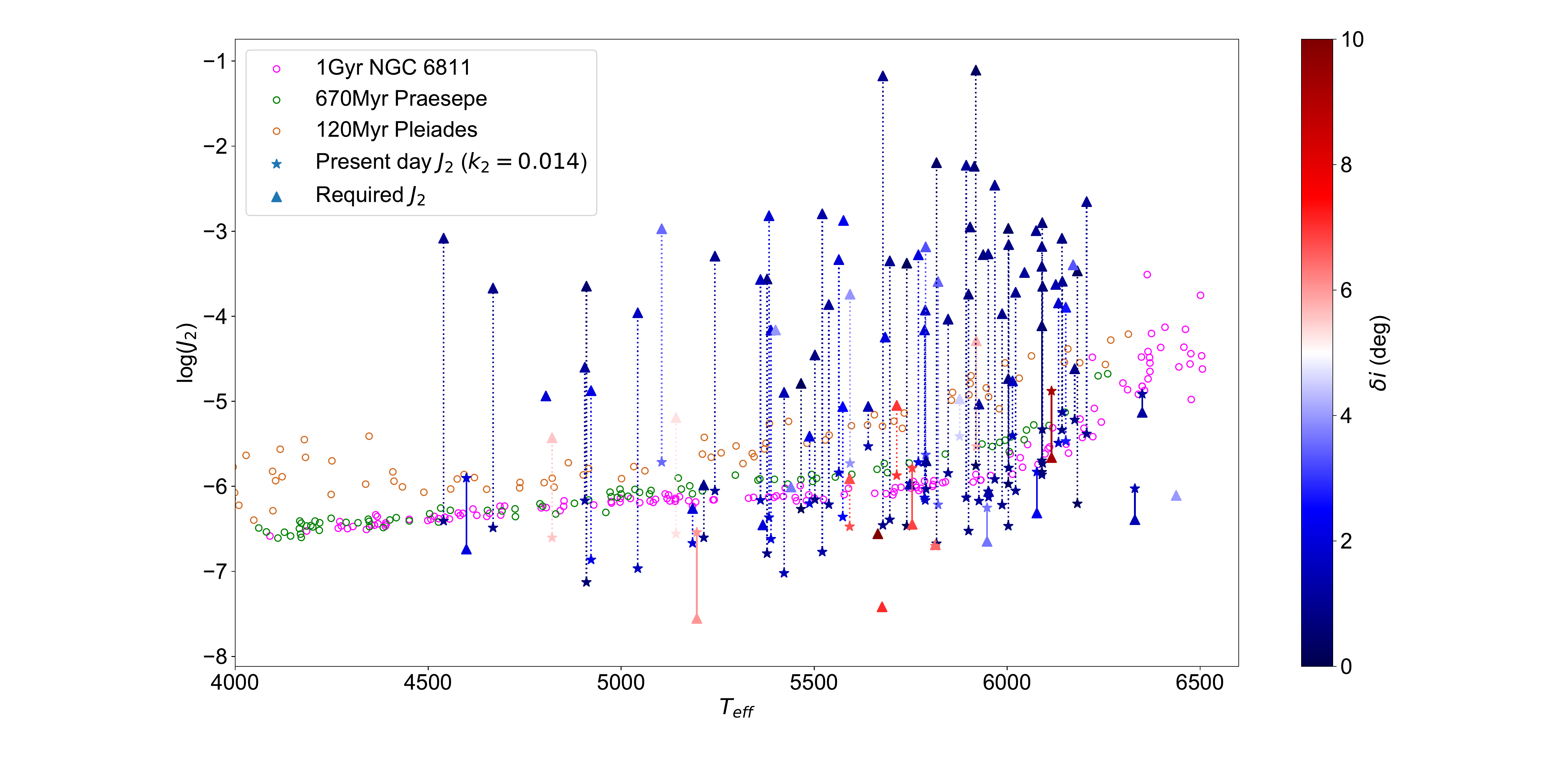}
 \caption{ The estimated $J_2$ of stars in three young clusters: Pleiades \citep{Rebull}, Praesepe \citep{Douglas} and NGC 6811 \citep{Curtis}. Also plotted is the present day $J_2$ (stars) and the $J_2$ required to induce large mutual inclinations (triangles) of the planet hosts in our sample. We linked the present day $J_2$ and required $J_2$ for each star with a solid line if the present day $J_2$ is greater than the required $J_2$ i.e. the star has enough quadrupole moment to induce large mutual inclination. Conversely, we linked the points with a dotted line indicating that the stars have to be rotating faster. The colorbar encodes the observed mutual inclination $\delta i$ of the innermost planet pair in each system. Although most of the current $J_2$ of the stars are smaller than the required $J_2$ to excite large mutual inclinations, the required $J_2$ are close to their current values and are smaller than those in clusters with younger ages ($\lesssim 1$Gyr) for the sample of close-in mutually-inclined systems ($a/R_\star<$ 5; $\delta i>$5$^\circ$). Thus, stellar obliquity is most likely able to excite the mutual inclination as long as the planets migrate to their current location within $\lesssim 1$Gyr.}
    \label{f:j2}
\end{figure*}

In Figure \ref{f:j2}, we plotted the estimated $J_2$ in the clusters and the sample of shortest-planet hosts. One can see the decay of $J_2$ as a function of age due to magnetic breaking. We also plotted the required $J_2$ that is capable of generating large mutual inclination in each system (as discussed in section \ref{s:obs}). We color code the plot using the observed mutual inclination obtained in \citet{Dai18}. 

Figure \ref{f:j2} shows that the required $J_2$ are mostly higher than the current ones, although the differences are small for the large mutual inclination systems. It suggests that it is challenging to enhance the mutual inclination with the current stellar oblateness. However, for systems with higher mutual inclinations, the required $J_2$ are lower than the $J_2$ of stars in the clusters ($10^{-5\sim-6}$). The differences in the $J_2$ are mainly due to their rotational rates. Thus, we expect the observed stars also had higher rotation rates similar to those in the clusters, and $J_2$ close to $10^{-5\sim-6}$ when they were younger ($\sim100$Myr$-1$Gyr). Therefore, for systems with high mutual inclinations, the oblateness of the stars should be high enough to excite the mutual inclinations, if the planets migrate to their current location within $\sim 1$Gyr. 

\section{Comparison with Observational Results}
\label{s:obs}
Mutual inclinations can be enhanced due to orbital precession when the stellar $J_2$ dominates over planetary coupling (as shown in section \ref{s:examples}). To verify whether the $J_2$ precession could indeed explain the observed mutual inclination signatures over innermost planet semi-major axes and planetar period ratios, we simulate the mutual inclination of the USPs and compare that with the observation. We first calculate the required $J_2$ in order to excite the mutual inclination of the planets. The mutual inclination obtained depends on stellar obliquity. Thus, we next estimate the possible mutual inclination due to stellar $J_2$ precession assuming different stellar obliquity distributions. We include all the observed planetary system with mutual inclination measured in \citet{Dai18} in our study.

\begin{figure*}[h]
\center
    \includegraphics[width=0.9\textwidth]{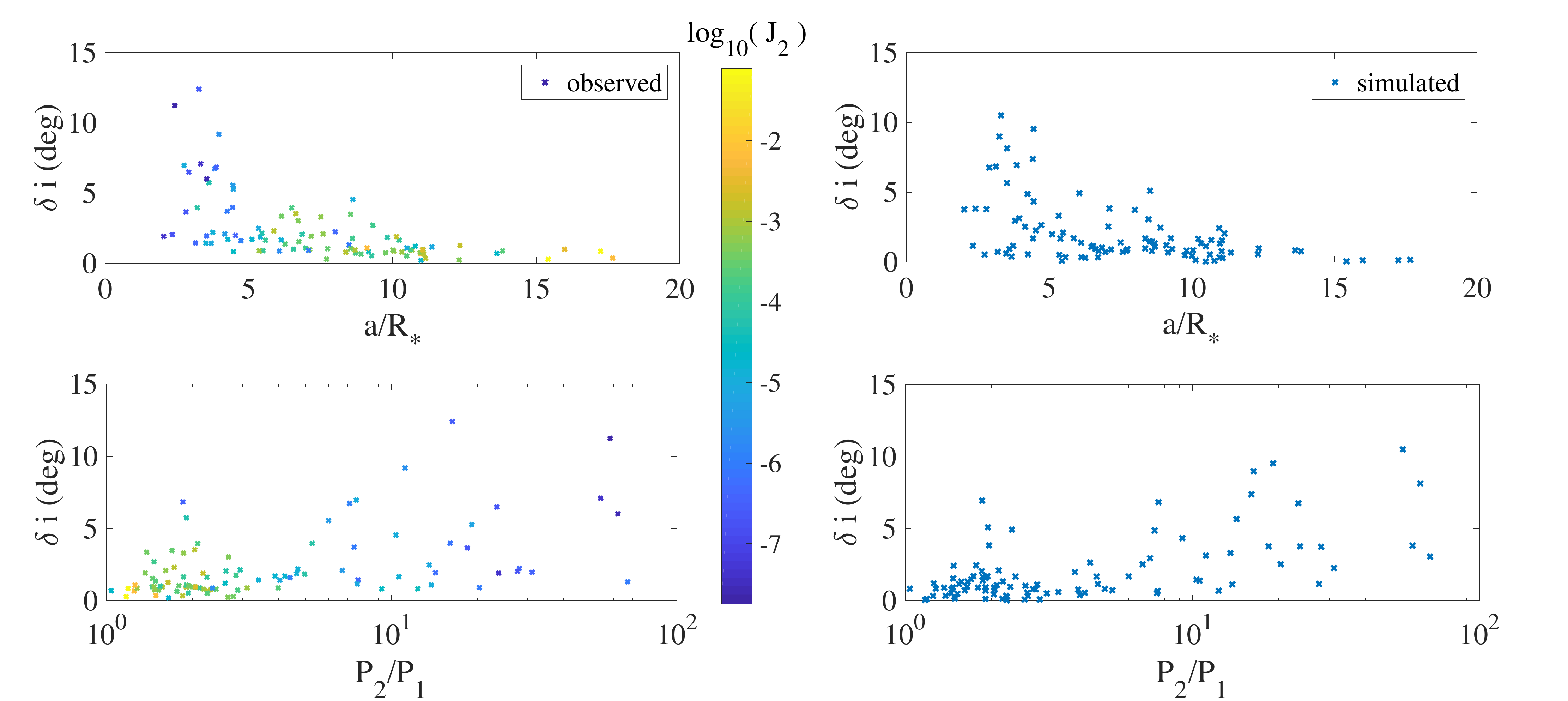}    
    \caption{Left panels: observed mutual inclination vs. semi-major axis of the innermost planet (top) and vs. the period ratio (bottom). The colors represent the required $J_2$ to increase the mutual inclinations. Right panels: simulated planetary mutual inclination as a function of the semi-major axis of the innermost planets (top) and the period ratio (bottom). We assume $J_2$ to be $3\times 10^{-6}$ for the simulated results, similar to the $J_2$ of stars in clusters with ages $\sim 100$Myr $-1$Gyr (see section \ref{s:emp}). The left panels show that the required $J_2$ to excite mutual inclination is small $\lesssim 10^{-5}$ when $a/R_*\lesssim5$ or $P_2/P_1\gtrsim5$, and the right panels show that the simulated mutual inclinations agree well with the observation (the two dimensional Kolmogorov-Smirnov test gives $p_{ks}=0.82$ between the upper panels and $0.9$ between the lower panels).}
    \label{f:obs}
\end{figure*}

To obtain the required $J_2$, we first calculated the slowest modal inclination frequencies for the observed planetary systems following the Lagrange-Laplace approach. We require the modes affect the innermost planets non-negligibly with eigenvector components larger than $0.1$, as discussed in section \ref{s:examples}. We obtain the masses of the planets following the empirical fit of the mass-radius relation \citep{Chen17}. Then, we obtain the $J_2$ that could lead to orbital precession frequency the same as the slowest modal frequency. As shown in section \ref{s:examples}, this is roughly the required $J_2$ which could enhance the mutual inclination between the planets. We present our results in the left panels of figure \ref{f:obs}.

The required $J_2$ is color coded in the plane of $\delta i$ vs. $a/R_*$ and $\delta i$ vs. $a/R_*$ in Figure \ref{f:obs}. It shows the required $J_2$ depends sensitively on $a/R_*$ and $P_2/P_1$. This is expected, since precession due to the stellar oblateness is stronger when the planet is closer to the host star (with small $a/R_*$) and when the planetary coupling is weaker (large $P_2/P_1$). In addition, it shows that it typically only requires a small $J_2 \sim 10^{-5\sim -6}$ to compete with the planetary coupling and to enhance the mutual inclination when $a/R_*<5$ or $P_2/P_1>5$.  

Next, we roughly estimate the mutual inclination using the required $J_2$. The maximum mutual inclination excited depends on the stellar obliquity ($\beta$), and we assume the stellar obliquity follows the von Mises Fisher distribution with $p=3$ (Fisher distribution) \citep{Fabrycky09}:
\begin{equation}
f(\beta)= \frac{\kappa}{2\sinh{\kappa}}e^{\kappa\cos{\beta}}\sin{\beta} ,
\end{equation}
where a larger $\kappa$ corresponds to a narrower distribution.

Based on 275 California-{\it Kepler} Survey targets which has rotation period, stellar radii and $V\sin(i)$ measured, \citet{Munoz18} obtained the statistical distribution of the stellar obliquity. They found that the distribution follows the Fisher distribution with $\kappa = 14.5^{+13.5}_{-6}$, where $\kappa=14.5$ corresponds to an average obliquity of $19^\circ$ and standard deviation of $10^\circ$. This distribution does not depend on planetary multiplicity, stellar multiplicity or stellar age statistically significantly, but $\kappa$ decreases (obliquity larger) when the planets are less massive, stellar metallicity are higher or planetary periods are longer. Selecting only the 118 targets in TEPCat Catalog with Rossiter-McLaughlin (RM) measurements (mostly hot jupiters), \citet{Munoz18} found that the obliquities follow a much broader distribution. It is not clear how the obliquities of the USP hosts distribute, and the $V\sin{i}$ method is prone to systematic errors. Thus, we try several distributions to simulate the mutual inclinations. 

USPs most likely migrated to their current locations when the systems were much younger ($\lesssim 1$Gyr) \citep[e.g.,][]{Lee17}. Thus, the mutual inclination could be excited when the system were younger and with a larger stellar $J_2$ potential. Therefore, we set stellar $J_2$ to be $J_{2,o}=3\times10^{-6}$ to simulate the mutual inclination according to the results of the clusters with age $\sim 1$Gyr, as discussed in section \ref{s:emp}. 

The mutual inclination can be obtained using the following heuristic method: 1: when the required $J_{2, req}$ is larger than $J_{2, o}$, planetary coupling dominates and the planetary system should be near co-planar. We set the mutual inclination to follow a normal distribution with standard deviation of $1.5^\circ$. The standard deviation is chosen so that the mutual inclination fits well in the region when the innermost planets are farther from the host star $a/R_* \gtrsim 10$. The general trend of the mutual inclination on $a/R_*$ or the period ratio does not depend on this assumption; 2: when $J_{2, req}<J_{2, o}<10J_{2, req}$, the maximum $\delta i$ increases with $J_{2, o}$ (e.g., figure \ref{f:frat}), so we set the mutual inclination to be $1+J_{2, o}/(10 J_{2, req}) \beta$, where $\beta$ is the stellar obliquity. 3: when $J_{2, o}>10J_{2, req}$, we set the mutual inclination to be $2\beta$, since the maximum mutual inclination due to stellar oblateness is around twice stellar obliquity as shown in section \ref{s:examples}. 

Briefly
\begin{equation}
  \delta i_{sim}  = \begin{cases}
{\rm follows~~} \mathcal{N}(0, 1.5) &\text{if $J_{2, o}<J_{2, req}$}\\
(1+J_{2, o}/(10 J_{2, req})) \beta &\text{else if $J_{2, o}<10 J_{2, req}$} \\
2\beta &\text{else} ,
\end{cases}
\end{equation}
where $\beta$ follows the von Mises-Fisher distribution ($p=3$) as described above. For illustration, we set $\kappa = 500$, so that the average stellar obliquity is around $3^\circ$. We note that this is on the low end of the possible obliquities as described above.

The right panels of Figure \ref{f:obs} illustrate the simulated results as a function of the distance of the host star $a/R_*$ (top right) as well as the period ratio ($P_2/P_1$, lower right). We can see that the simulated mutual inclinations also show the dependence on the innermost planetary semi-major axis and planetary period ratio, agreeing well with that of the observation. Using a very narrow distribution ($\kappa=500$), the 2-dimensional ks test in the plane of $\delta i - a/R_*$ and the plane of $\delta i - P_2/P_1$ both yield large p-values ($p_{ks}=0.82$ and $p_{ks}=0.9$ separately). This shows that the simulated mutual inclination can agree well with the observation. 

Note that the observed mutual inclination is only the minimum inclination, since \citet{Dai18} only took the difference in the line of sight inclinations of the innermost two planets as an proxy for the mutual inclination, and the obtained mutual inclination neglects the possibilities of non-transiting planets with larger inclinations. Thus, the mutual inclination may actually agree well with the observation with a wider obliquity distribution ($\kappa < 500$). For instance, the maximum simulated inclination reaches $70^\circ$ with $\kappa = 15$ as indicated by the CKS survey \citep{Munoz18}. Using a narrower distribution of $\kappa=150$ (average obliquity of $6^\circ$, and standard deviation of $3^\circ$), the maximum simulated inclination reaches $20^\circ$. The general trends on the innermost planetary semi-major axis as well as the planetary period ratio remain independent of the value of $\kappa$. This supports that the larger mutual inclination can be obtained due to stellar obliquity alone, and one does not need to resort to migration processes that excite planetary inclinations.

\section{Discussions}
\label{s:dis}

In this work, we find that the trend identified in \citealt{Dai18} that planets with shorter orbital periods have a higher range in mutual inclinations can be explained in part due to nodal precession caused by stellar oblateness. This mechanism also explains the dependence of observed mutual inclination with innermost planet semi-major axis and planet-planet orbital period ratio.

Thus, it is possible that the USPs migrate to their current locations with low mutual inclinations, and then their orbits precess due to the stellar $J_2$ potential at different rates to excite the mutual inclination. USP formation mechanisms, such as disk migration and tidal decay \citep{Schlaufman10}, collisions and mergers prior to migration \citep{Terquem14} and in-situ formation and tidal decay \citep{Lee17} are still compatible with the observation of the high mutual inclination of USPs.

With an assumed stellar obliquity distribution following the von Mises Fisher with $\kappa = 500$ (average obliquity $\langle \delta \beta \rangle \sim 3^\circ$), we find that the simulated mutual inclination matches well with that of the observation. However, the observed mutual inclination reported in \citet{Dai18} is the minimum inclination. In addition, there could be non-transiting planetary companions with higher mutual inclination. This suggests that the theoretical model with $\kappa=500$ could produce lower mutual inclination than the true mutual inclination of the systems. It is possible that the stellar spin-orbit misalignment is higher than the Fisher distribution with $\kappa = 500$. We note that using a von Mises Fisher distribution with $\kappa = 150$ (standard deviation of the obliquity $\sigma_\beta \sim 3^\circ$), the maximum mutual inclination increase up to $\sim 20^\circ$, slightly higher than the observed minimum mutual inclination. 

Dynamical migration mechanisms due to planetary interactions may also contribute to the large mutual inclinations of USPs close to the host stars \citep{Petrovich19, Pu19}. The $J_2$ precession timescales of the USPs are around $10^4$yrs, and thus, the transit duration variations are too small to be detected to distinguish the mechanisms. On the other hand, one could use the stellar obliquity distribution of the USPs. Observations of stellar obliquities with small planets $\sim 2R_{\oplus}$ will be available in the near future \citep{Johnson19}. A better understanding on the stellar obliquity distribution of the USP hosts will help determine the mutual inclination distribution due to stellar $J_2$ precession, and to understand the role of near coplanar disk migration and migration due to planet-planet secular interactions in USP planet formation.

\begin{longtable*}{llllll}
\tablehead{
\colhead{Name} & \colhead{$a/R_*$} & \colhead{$\delta i$ (deg)} & \colhead{$P_{rot}$ (days)} &\colhead{$J_{2, today}$} & \colhead{$J_{2, req}$}
}
EPIC-211305568 & $2.42^{+0.02}_{-0.02}$ & $11.24^{+4.32}_{-7.27}$ & $41.90^{+0.10}_{-0.10}$ & $7.65^{+0.35}_{-0.35} \times 10^{-9}$ &$1.79 \times 10^{-8}$ \\ 
Kepler-312 & $3.95^{+0.11}_{-0.11}$ & $9.19^{+1.15}_{-1.29}$ & $6.10^{+0.10}_{-0.20}$ & $6.16^{+1.09}_{-0.75} \times 10^{-6}$ &$2.18 \times 10^{-6}$ \\ 
KOI-787.0 & $2.74^{+0.11}_{-0.10}$ & $6.97^{+4.51}_{-5.04}$ & $12.40^{+0.80}_{-0.50}$ & $6.25^{+1.63}_{-1.52} \times 10^{-7}$ &$8.96 \times 10^{-6}$ \\ 
Kepler-1047 & $3.87^{+0.13}_{-0.11}$ & $6.84^{+3.22}_{-3.18}$ & $18.20^{+0.10}_{-0.20}$ & $7.62^{+1.06}_{-0.91} \times 10^{-7}$ &$3.59 \times 10^{-7}$ \\ 
Kepler-1067 & $3.81^{+0.08}_{-0.08}$ & $6.74^{+1.75}_{-2.69}$ & $18.00^{+0.20}_{-0.20}$ & $1.57^{+0.17}_{-0.14} \times 10^{-7}$ &$1.23 \times 10^{-6}$ \\ 
K2-106 & $2.91^{+0.06}_{-0.06}$ & $6.49^{+3.93}_{-5.02}$ & $25.80^{+3.40}_{-3.10}$ & $9.27^{+3.65}_{-2.55} \times 10^{-8}$ &$2.06 \times 10^{-7}$ \\ 
Kepler-607 & $3.53^{+0.08}_{-0.07}$ & $6.02^{+2.93}_{-4.11}$ & $18.50^{+0.10}_{-0.20}$ & $1.34^{+0.12}_{-0.12} \times 10^{-7}$ &$2.80 \times 10^{-8}$ \\ 
Kepler-207 & $3.61^{+0.09}_{-0.10}$ & $5.75^{+2.10}_{-2.30}$ & $13.90^{+0.10}_{-0.30}$ & $1.38^{+0.17}_{-0.12} \times 10^{-6}$ &$5.15 \times 10^{-5}$ \\ 
KOI-2393.0 & $4.44^{+0.08}_{-0.08}$ & $5.56^{+1.30}_{-1.99}$ & $16.90^{+0.20}_{-0.10}$ & $1.16^{+0.12}_{-0.12} \times 10^{-7}$ &$3.73 \times 10^{-6}$ \\ 
KOI-1360.0 & $4.46^{+0.08}_{-0.08}$ & $5.27^{+2.25}_{-3.22}$ & $15.80^{+0.10}_{-0.20}$ & $1.29^{+0.16}_{-0.12} \times 10^{-7}$ &$6.44 \times 10^{-6}$ \\ 
Kepler-335 & $8.61^{+0.23}_{-0.22}$ & $4.55^{+0.32}_{-0.31}$ & $13.20^{+1.50}_{-1.00}$ & $1.81^{+0.57}_{-0.51} \times 10^{-6}$ &$1.07 \times 10^{-5}$ \\ 
Kepler-363 & $6.49^{+0.16}_{-0.18}$ & $3.97^{+1.38}_{-0.86}$ & $17.30^{+0.20}_{-0.50}$ & $8.67^{+1.26}_{-1.26} \times 10^{-7}$ &$1.82 \times 10^{-4}$ \\ 
Kepler-990 & $2.81^{+0.12}_{-0.13}$ & $3.66^{+3.88}_{-2.74}$ & $16.00^{+0.30}_{-0.20}$ & $2.61^{+0.38}_{-0.38} \times 10^{-7}$ &$2.25 \times 10^{-7}$ \\ 
Kepler-326 & $6.63^{+0.15}_{-0.17}$ & $3.54^{+1.41}_{-1.51}$ & $9.60^{+0.10}_{-0.10}$ & $8.98^{+1.10}_{-1.01} \times 10^{-7}$ &$1.07 \times 10^{-3}$ \\ 
Kepler-203 & $8.54^{+0.27}_{-0.24}$ & $3.49^{+1.06}_{-1.34}$ & $16.30^{+0.10}_{-0.20}$ & $2.84^{+0.38}_{-0.31} \times 10^{-7}$ &$2.56 \times 10^{-4}$ \\ 
Kepler-625 & $7.50^{+0.23}_{-0.19}$ & $3.31^{+1.07}_{-1.30}$ & $12.30^{+0.10}_{-0.10}$ & $1.10^{+0.13}_{-0.11} \times 10^{-6}$ &$6.54 \times 10^{-4}$ \\ 
Kepler-381 & $9.31^{+0.28}_{-0.30}$ & $2.71^{+1.37}_{-1.76}$ & $10.20^{+0.10}_{-0.20}$ & $1.58^{+0.18}_{-0.13} \times 10^{-6}$ &$1.28 \times 10^{-4}$ \\ 
Kepler-198 & $5.34^{+0.16}_{-0.12}$ & $2.49^{+1.35}_{-1.96}$ & $15.70^{+1.40}_{-1.90}$ & $2.04^{+0.79}_{-0.42} \times 10^{-7}$ &$8.67 \times 10^{-6}$ \\ 
Kepler-140 & $8.01^{+0.24}_{-0.22}$ & $2.25^{+0.68}_{-1.02}$ & $12.00^{+0.20}_{-0.10}$ & $6.90^{+0.72}_{-0.77} \times 10^{-7}$ &$4.86 \times 10^{-7}$ \\ 
KOI-2250.0 & $3.74^{+0.08}_{-0.08}$ & $2.20^{+1.52}_{-1.49}$ & $24.20^{+0.40}_{-0.30}$ & $6.39^{+0.83}_{-0.78} \times 10^{-8}$ &$1.33 \times 10^{-5}$ \\ 
Kepler-1322 & $4.16^{+0.18}_{-0.17}$ & $2.10^{+1.80}_{-1.51}$ & $14.40^{+0.20}_{-0.10}$ & $2.93^{+0.70}_{-0.55} \times 10^{-7}$ &$3.89 \times 10^{-6}$ \\ 
Kepler-1365 & $7.58^{+0.21}_{-0.19}$ & $2.10^{+1.61}_{-0.96}$ & $17.50^{+0.20}_{-0.20}$ & $8.94^{+1.02}_{-0.99} \times 10^{-7}$ &$5.29 \times 10^{-4}$ \\ 
Kepler-322 & $6.85^{+0.13}_{-0.14}$ & $2.07^{+1.03}_{-1.19}$ & $19.10^{+0.30}_{-0.20}$ & $1.12^{+0.08}_{-0.08} \times 10^{-7}$ &$6.89 \times 10^{-5}$ \\ 
K2-141 & $2.33^{+0.03}_{-0.03}$ & $2.05^{+2.25}_{-1.44}$ & $7.00^{+0.30}_{-0.30}$ & $5.84^{+0.89}_{-0.74} \times 10^{-7}$ &$1.83 \times 10^{-7}$ \\ 
K2-229 & $3.53^{+0.07}_{-0.06}$ & $1.95^{+2.46}_{-1.43}$ & $19.30^{+1.90}_{-2.60}$ & $1.00^{+0.42}_{-0.22} \times 10^{-7}$ &$5.49 \times 10^{-7}$ \\ 
Kepler-1542 & $7.17^{+0.15}_{-0.15}$ & $1.93^{+1.61}_{-1.41}$ & $13.10^{+0.30}_{-0.10}$ & $6.75^{+0.57}_{-0.79} \times 10^{-7}$ &$4.64 \times 10^{-4}$ \\ 
KOI-1843.0 & $2.03^{+0.02}_{-0.02}$ & $1.92^{+2.53}_{-1.41}$ & $34.40^{+0.50}_{-0.40}$ & $1.53^{+0.10}_{-0.11} \times 10^{-8}$ &$4.49 \times 10^{-8}$ \\ 
Kepler-18 & $10.14^{+0.24}_{-0.23}$ & $1.90^{+0.13}_{-0.13}$ & $16.30^{+0.40}_{-0.20}$ & $2.01^{+0.23}_{-0.24} \times 10^{-7}$ &$1.52 \times 10^{-3}$ \\ 
Kepler-219 & $9.82^{+0.29}_{-0.26}$ & $1.85^{+0.36}_{-0.43}$ & $17.30^{+1.70}_{-1.80}$ & $3.42^{+1.22}_{-0.83} \times 10^{-7}$ &$6.91 \times 10^{-5}$ \\ 
Kepler-356 & $8.60^{+0.25}_{-0.24}$ & $1.77^{+0.33}_{-0.29}$ & $10.40^{+0.10}_{-0.10}$ & $1.51^{+0.20}_{-0.15} \times 10^{-6}$ &$1.44 \times 10^{-4}$ \\ 
Kepler-524 & $4.26^{+0.09}_{-0.10}$ & $1.71^{+1.78}_{-1.21}$ & $11.80^{+0.10}_{-0.10}$ & $1.83^{+0.21}_{-0.20} \times 10^{-6}$ &$1.74 \times 10^{-5}$ \\ 
Kepler-32 & $5.10^{+0.05}_{-0.05}$ & $1.70^{+1.43}_{-1.23}$ & $38.00^{+0.70}_{-0.70}$ & $1.41^{+0.13}_{-0.12} \times 10^{-8}$ &$1.55 \times 10^{-5}$ \\ 
Kepler-732 & $6.12^{+0.05}_{-0.06}$ & $1.67^{+0.59}_{-0.73}$ & $34.90^{+0.70}_{-1.50}$ & $4.57^{+1.19}_{-0.80} \times 10^{-8}$ &$1.16 \times 10^{-5}$ \\ 
Kepler-100 & $10.23^{+0.23}_{-0.18}$ & $1.66^{+0.43}_{-0.35}$ & $25.90^{+6.00}_{-2.40}$ & $3.19^{+0.97}_{-1.26} \times 10^{-7}$ &$1.18 \times 10^{-4}$ \\ 
Kepler-755 & $5.58^{+0.11}_{-0.11}$ & $1.64^{+2.83}_{-1.11}$ & $29.81^{+0.24}_{-0.24}$ & $5.05^{+0.38}_{-0.41} \times 10^{-8}$ &$1.10 \times 10^{-4}$ \\ 
Kepler-1311 & $4.72^{+0.08}_{-0.09}$ & $1.61^{+1.67}_{-1.16}$ & $24.60^{+0.40}_{-0.30}$ & $4.99^{+0.59}_{-0.71} \times 10^{-7}$ &$1.06 \times 10^{-6}$ \\ 
Kepler-1371 & $6.74^{+0.12}_{-0.11}$ & $1.54^{+1.72}_{-1.09}$ & $14.40^{+0.10}_{-0.20}$ & $3.20^{+0.31}_{-0.26} \times 10^{-7}$ &$2.71 \times 10^{-4}$ \\ 
Kepler-1340 & $3.14^{+0.08}_{-0.08}$ & $1.45^{+1.60}_{-1.02}$ & $12.60^{+0.90}_{-0.40}$ & $4.39^{+0.95}_{-0.99} \times 10^{-7}$ &$4.06 \times 10^{-7}$ \\ 
KOI-191.0 & $3.50^{+0.07}_{-0.07}$ & $1.44^{+1.07}_{-0.89}$ & $36.20^{+0.20}_{-0.30}$ & $4.44^{+0.67}_{-0.58} \times 10^{-8}$ &$1.27 \times 10^{-5}$ \\ 
KOI-1239.0 & $3.69^{+0.10}_{-0.10}$ & $1.43^{+1.34}_{-1.00}$ & $6.70^{+1.50}_{-1.60}$ & $1.38^{+1.41}_{-0.59} \times 10^{-6}$ &$8.71 \times 10^{-6}$ \\ 
Kepler-111 & $8.48^{+0.17}_{-0.23}$ & $1.32^{+0.67}_{-0.73}$ & $16.20^{+0.20}_{-0.10}$ & $3.48^{+0.33}_{-0.35} \times 10^{-7}$ &$8.69 \times 10^{-7}$ \\ 
Kepler-783 & $12.35^{+0.32}_{-0.30}$ & $1.28^{+1.39}_{-0.90}$ & $23.90^{+1.40}_{-3.30}$ & $7.90^{+3.50}_{-1.30} \times 10^{-8}$ &$1.60 \times 10^{-3}$ \\ 
Kepler-411 & $10.79^{+0.18}_{-0.12}$ & $1.22^{+0.15}_{-0.16}$ & $10.40^{+0.20}_{-0.10}$ & $3.17^{+0.23}_{-0.24} \times 10^{-7}$ &$2.52 \times 10^{-5}$ \\ 
Kepler-853 & $11.37^{+0.30}_{-0.27}$ & $1.17^{+0.60}_{-0.76}$ & $5.30^{+0.40}_{-0.30}$ & $5.67^{+1.19}_{-1.14} \times 10^{-6}$ &$7.43 \times 10^{-6}$ \\ 
Kepler-466 & $10.50^{+0.32}_{-0.44}$ & $1.10^{+0.78}_{-0.74}$ & $14.90^{+0.20}_{-0.20}$ & $3.14^{+0.91}_{-0.65} \times 10^{-7}$ &$9.28 \times 10^{-6}$ \\ 
Kepler-526 & $9.11^{+0.24}_{-0.20}$ & $1.09^{+1.23}_{-0.80}$ & $24.40^{+0.20}_{-0.30}$ & $3.43^{+0.56}_{-0.32} \times 10^{-7}$ &$5.98 \times 10^{-3}$ \\ 
Kepler-406 & $6.98^{+0.13}_{-0.14}$ & $1.09^{+0.95}_{-0.74}$ & $16.80^{+0.20}_{-0.20}$ & $2.85^{+0.32}_{-0.29} \times 10^{-7}$ &$1.37 \times 10^{-4}$ \\ 
Kepler-1271 & $7.74^{+0.21}_{-0.23}$ & $1.05^{+1.19}_{-0.76}$ & $5.30^{+0.40}_{-0.30}$ & $3.48^{+0.96}_{-0.83} \times 10^{-6}$ &$2.58 \times 10^{-4}$ \\ 
Kepler-107 & $6.55^{+0.15}_{-0.14}$ & $1.01^{+0.93}_{-0.68}$ & $16.20^{+4.30}_{-3.90}$ & $6.66^{+6.34}_{-2.81} \times 10^{-7}$ &$9.26 \times 10^{-5}$ \\ 
Kepler-277 & $15.99^{+0.12}_{-0.18}$ & $1.00^{+0.67}_{-0.75}$ & $22.60^{+0.30}_{-0.20}$ & $5.65^{+0.75}_{-0.70} \times 10^{-7}$ &$3.48 \times 10^{-3}$ \\ 
Kepler-338 & $11.06^{+0.27}_{-0.23}$ & $0.99^{+0.57}_{-0.67}$ & $27.81^{+6.47}_{-22.03}$ & $4.04^{+98.23}_{-1.62} \times 10^{-7}$ &$5.42 \times 10^{-4}$ \\ 
Kepler-213 & $7.11^{+0.22}_{-0.18}$ & $0.98^{+0.32}_{-0.39}$ & $22.10^{+0.10}_{-0.30}$ & $1.89^{+0.23}_{-0.17} \times 10^{-7}$ &$4.48 \times 10^{-4}$ \\ 
Kepler-1581 & $10.63^{+0.27}_{-0.27}$ & $0.97^{+0.89}_{-0.65}$ & $19.40^{+0.20}_{-0.40}$ & $4.14^{+0.57}_{-0.42} \times 10^{-7}$ &$1.91 \times 10^{-4}$ \\ 
Kepler-450 & $11.05^{+0.20}_{-0.17}$ & $0.97^{+0.43}_{-0.40}$ & $10.40^{+0.10}_{-0.10}$ & $1.93^{+0.18}_{-0.17} \times 10^{-6}$ &$2.23 \times 10^{-3}$ \\ 
Kepler-221 & $10.01^{+0.20}_{-0.18}$ & $0.95^{+0.50}_{-0.52}$ & $9.40^{+0.20}_{-0.10}$ & $4.13^{+0.30}_{-0.34} \times 10^{-7}$ &$5.09 \times 10^{-4}$ \\ 
Kepler-135 & $10.68^{+0.30}_{-0.29}$ & $0.94^{+0.91}_{-0.68}$ & $16.10^{+0.20}_{-0.20}$ & $6.38^{+0.82}_{-0.79} \times 10^{-7}$ &$6.62 \times 10^{-4}$ \\ 
Kepler-969 & $7.07^{+0.11}_{-0.11}$ & $0.92^{+0.87}_{-0.69}$ & $18.00^{+0.30}_{-0.20}$ & $1.16^{+0.09}_{-0.08} \times 10^{-7}$ &$1.04 \times 10^{-6}$ \\ 
Kepler-323 & $5.49^{+0.08}_{-0.11}$ & $0.91^{+0.94}_{-0.63}$ & $17.40^{+0.20}_{-0.10}$ & $2.81^{+0.32}_{-0.32} \times 10^{-7}$ &$1.07 \times 10^{-4}$ \\ 
Kepler-80 & $5.36^{+0.05}_{-0.07}$ & $0.90^{+0.95}_{-0.62}$ & $12.90^{+0.10}_{-0.10}$ & $1.83^{+0.14}_{-0.13} \times 10^{-7}$ &$8.29 \times 10^{-4}$ \\ 
Kepler-202 & $13.83^{+0.19}_{-0.19}$ & $0.89^{+0.15}_{-0.15}$ & $14.00^{+6.50}_{-6.00}$ & $1.52^{+3.42}_{-0.85} \times 10^{-7}$ &$2.14 \times 10^{-4}$ \\ 
Kepler-142 & $6.06^{+0.11}_{-0.14}$ & $0.87^{+0.76}_{-0.62}$ & $14.60^{+0.20}_{-0.30}$ & $4.32^{+0.51}_{-0.42} \times 10^{-7}$ &$2.00 \times 10^{-6}$ \\ 
Kepler-431 & $10.05^{+0.24}_{-0.24}$ & $0.87^{+1.07}_{-0.63}$ & $17.60^{+0.20}_{-0.20}$ & $7.69^{+0.74}_{-0.73} \times 10^{-7}$ &$6.95 \times 10^{-4}$ \\ 
Kepler-200 & $17.24^{+0.45}_{-0.50}$ & $0.86^{+0.55}_{-0.43}$ & $19.50^{+0.20}_{-0.20}$ & $1.63^{+0.19}_{-0.17} \times 10^{-7}$ &$6.71 \times 10^{-2}$ \\ 
Kepler-218 & $9.17^{+0.18}_{-0.18}$ & $0.84^{+0.74}_{-0.62}$ & $15.60^{+0.10}_{-0.10}$ & $3.26^{+0.34}_{-0.32} \times 10^{-7}$ &$3.51 \times 10^{-5}$ \\ 
Kepler-342 & $4.46^{+0.08}_{-0.11}$ & $0.84^{+1.17}_{-0.60}$ & $7.10^{+0.40}_{-0.20}$ & $2.82^{+0.42}_{-0.49} \times 10^{-6}$ &$2.42 \times 10^{-5}$ \\ 
Kepler-116 & $10.32^{+0.29}_{-0.27}$ & $0.81^{+1.68}_{-0.69}$ & $8.30^{+0.80}_{-0.60}$ & $2.14^{+0.61}_{-0.53} \times 10^{-6}$ &$8.27 \times 10^{-4}$ \\ 
Kepler-314 & $8.37^{+0.17}_{-0.19}$ & $0.81^{+0.75}_{-0.56}$ & $25.60^{+0.40}_{-0.50}$ & $7.58^{+0.84}_{-0.66} \times 10^{-8}$ &$2.75 \times 10^{-4}$ \\ 
Kepler-402 & $9.76^{+0.28}_{-0.25}$ & $0.75^{+0.74}_{-0.53}$ & $9.20^{+0.20}_{-0.40}$ & $9.36^{+1.75}_{-1.12} \times 10^{-7}$ &$3.86 \times 10^{-4}$ \\ 
Kepler-376 & $8.71^{+0.24}_{-0.26}$ & $0.74^{+0.98}_{-0.54}$ & $36.90^{+0.40}_{-0.60}$ & $1.40^{+0.18}_{-0.16} \times 10^{-7}$ &$1.82 \times 10^{-4}$ \\ 
Kepler-132 & $13.63^{+0.68}_{-0.67}$ & $0.70^{+0.66}_{-0.47}$ & $20.70^{+0.20}_{-0.20}$ & $1.59^{+0.46}_{-0.33} \times 10^{-7}$ &$1.85 \times 10^{-5}$ \\ 
Kepler-216 & $11.08^{+0.25}_{-0.30}$ & $0.68^{+0.59}_{-0.49}$ & $10.10^{+0.70}_{-0.90}$ & $2.18^{+0.73}_{-0.46} \times 10^{-6}$ &$1.26 \times 10^{-3}$ \\ 
Kepler-208 & $8.90^{+0.25}_{-0.22}$ & $0.66^{+0.69}_{-0.42}$ & $11.70^{+0.50}_{-0.30}$ & $8.70^{+1.31}_{-1.35} \times 10^{-7}$ &$2.25 \times 10^{-4}$ \\ 
Kepler-403 & $9.27^{+0.16}_{-0.18}$ & $0.54^{+0.60}_{-0.41}$ & $21.40^{+0.30}_{-0.30}$ & $6.86^{+1.18}_{-0.83} \times 10^{-7}$ &$7.73 \times 10^{-5}$ \\ 
Kepler-141 & $10.49^{+0.17}_{-0.17}$ & $0.52^{+0.55}_{-0.37}$ & $33.00^{+0.60}_{-0.60}$ & $3.47^{+0.34}_{-0.28} \times 10^{-8}$ &$2.26 \times 10^{-4}$ \\ 
Kepler-804 & $17.66^{+0.42}_{-0.45}$ & $0.37^{+0.36}_{-0.26}$ & $29.00^{+1.60}_{-1.50}$ & $9.89^{+1.99}_{-1.74} \times 10^{-8}$ &$6.40 \times 10^{-3}$ \\ 
Kepler-197 & $11.14^{+0.28}_{-0.28}$ & $0.36^{+0.45}_{-0.25}$ & $15.00^{+0.20}_{-0.10}$ & $4.98^{+0.54}_{-0.53} \times 10^{-7}$ &$1.08 \times 10^{-3}$ \\ 
Kepler-89 & $7.70^{+0.13}_{-0.10}$ & $0.29^{+0.37}_{-0.19}$ & $22.30^{+0.30}_{-0.30}$ & $2.91^{+0.35}_{-0.29} \times 10^{-7}$ &$3.42 \times 10^{-4}$ \\ 
Kepler-36 & $15.42^{+0.14}_{-0.20}$ & $0.29^{+0.28}_{-0.21}$ & $17.60^{+0.20}_{-0.30}$ & $8.14^{+0.99}_{-0.76} \times 10^{-7}$ &$7.84 \times 10^{-2}$ \\ 
Kepler-321 & $12.32^{+0.37}_{-0.32}$ & $0.25^{+0.21}_{-0.16}$ & $20.10^{+0.20}_{-0.20}$ & $1.60^{+0.16}_{-0.15} \times 10^{-7}$ &$4.21 \times 10^{-4}$ \\ 
Kepler-20 & $10.99^{+0.15}_{-0.14}$ & $0.21^{+0.34}_{-0.16}$ & $13.90^{+0.10}_{-0.30}$ & $2.53^{+0.29}_{-0.18} \times 10^{-7}$ &$1.62 \times 10^{-5}$ \\ 
\hline
\\
\caption{Planetary System Properties \label{tab:Prot}} 
\end{longtable*}

\section*{Acknowledgments}
The authors thank Josh Winn and the referee for helpful comments and suggestions on the manuscript.




\bibliographystyle{hapj}
\bibliography{ref.bib}

\end{document}